\documentclass[showpacs,twocolumn,preprintnumbers,amsmath,amssymb,prl,epsf,superscriptaddress]{revtex4-1}
\usepackage{graphicx}
\usepackage{nicefrac}
\usepackage[squaren]{SIunits}
\usepackage{color}
\begin{document}
\title{Macroscopic Hong-Ou-Mandel interference}
\author{Timur Sh.~Iskhakov}
\affiliation{Max Planck Institute for the Science of Light,
G\"unther-Scharowsky-Stra\ss{}e 1/Bau 24, 91058 Erlangen, Germany}
\author{Kirill~Yu.~Spasibko}
\affiliation{Department of Physics, M.V.Lomonosov Moscow State University, \\ Leninskie Gory, 119991 Moscow,
Russia}
\author{Maria~V.~Chekhova}
\affiliation{Max Planck Institute for the Science of Light, G\"unther-Scharowsky-Stra\ss{}e 1/Bau 24, 91058
Erlangen, Germany} \affiliation{Department of Physics, M.V.Lomonosov Moscow State University, \\ Leninskie Gory,
119991 Moscow, Russia}\affiliation{University of Erlangen-N\"urnberg, Staudtstrasse 7/B2, 91058 Erlangen, Germany}
\author{Gerd Leuchs}
\affiliation{Max Planck Institute for the Science of Light, G\"unther-Scharowsky-Stra\ss{}e 1/Bau 24, 91058
Erlangen, Germany} \affiliation{University of Erlangen-N\"urnberg, Staudtstrasse 7/B2, 91058 Erlangen, Germany}
\vspace{-10mm}

\begin{abstract}
We report on a Hong-Ou-Mandel (HOM) interference experiment for quantum states
with photon numbers per mode as large as $10^6$  generated via high-gain parametric
down conversion (PDC). The standard technique of coincidence counting leads in this case to a dip with a very low visibility. By measuring, instead of coincidence rate, the variance of the photon-number difference, we observe a peak with 99.99\% visibility. From the shape of the peak, one can infer
information about the temporal mode structure of the PDC radiation, including the degree of frequency/time entanglement.

\end{abstract}
\vspace{5mm}
\maketitle \narrowtext

One of the most fascinating manifestations of quantum interference is the Hong-Ou-Mandel (HOM) dip, an effect that started a whole new direction in quantum optics  First reported in 1987~\cite{Mandel,Alley}, it consists of the interference of two single-photon wavepackets on a 50\% beamsplitter (BS). When indistinguishable photons simultaneously arrive at different input ports of a BS, they always depart from the same port, due to the destructive interference of the probability amplitudes to be both reflected or both transmitted~\cite{constructive}. The effect is usually observed as a 'dip' in coincidences, its width determined by the coherence time of the photons and not affected by the group-velocity dispersion~\cite{ASteinberg1992}. HOM interference is widely used for identity tests of single-photon states~\cite{Yamamoto}, including ones generated by a quantum dot and a laser~\cite{Bennett}, conditionally prepared via parametric down-conversion (PDC) and a highly attenuated coherent state~\cite{Katiuscia}, produced via two consequent emissions of a single atom~\cite{Kuhn} or molecule~\cite{Zumbusch} or by two different molecules~\cite{Sandoghdar}. Other important applications are the  implementation of a photon Fock-state filter~\cite{FockFilter} and quantum-optical coherence tomography~\cite{QOCT}. It is worth mentioning that if, as in most cases, the photons are generated via PDC, the level of coincidence rate in the 'dip' is not zero but is given by the accidental coincidences  rate~\cite{Klyshko1994}. For this reason, the visibility of the dip reduces down to 33\% at high-gain PDC~\cite{DeMartini} and becomes vanishingly low if many modes are involved.


In this paper we expand the boundaries of the HOM effect applications and observe destructive HOM interference for bright twin beams (containing, on the average, up to $8\cdot10^5$ photons per mode) generated via high-gain PDC in a traveling-wave optical parametrical amplifier. Namely, instead of two single-photon states we sent to the beamsplitter two macroscopic states containing exactly equal number of quanta.
We suggest an alternative approach of observing the HOM effect, based on the measurement of the normalized variance of the difference signal at the output ports of the beamsplitter. This technique is robust against the multi-mode detection~\cite{JETP2008} and allows one to observe the interference with a high visibility.

\begin{figure}[htb]
\centering
\includegraphics[width=0.48\textwidth]{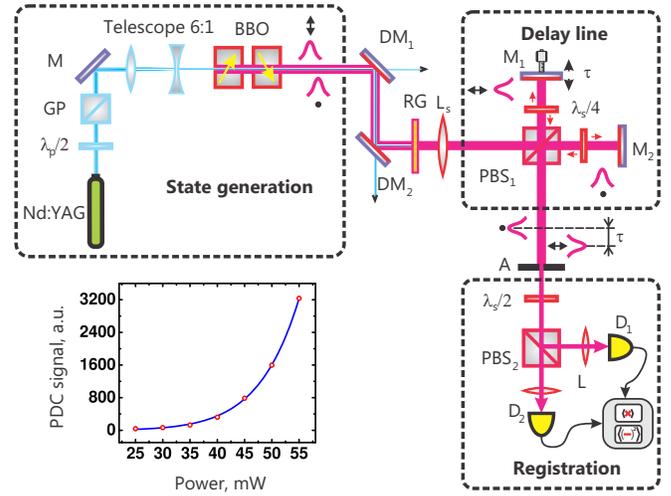}
\caption{(Color online) Experimental layout for the macroscopic HOM interference: Nd:YAG ($3\omega$), third harmonic of the Nd:YAG laser; $\lambda_p/2$, half-wave plate for 354.7 nm; GP, Glan prism; M, mirror; BBO, two 5 mm type-II beta barium borate crystals with the optical axes tilted oppositely with respect to the pump; $\mathrm {DM_{1,2}}$, dichroic mirrors with high transmission for the
pump and $99.5\%$ reflection for the down-converted radiation; RG, red glass filter $\mathrm{RG630}$ with AR coating at $709.3$~nm; $\mathrm {L_s}$, lens with the focal length 0.75 m; $\mathrm {PBS_{1,2}}$, polarization beamsplitter; $\lambda_s/4$, quarter-wave plate for 709.3 nm; $\mathrm {M_{1,2}}$, high-reflection mirrors at 709.3 nm; $\mathrm {A}$, 3 mm aperture placed in the focal plane of the lens $\mathrm {L_s}$; $\lambda_s/2$, half-wave plate for 709.3 nm; L, lens; $\mathrm {D_{1,2}}$, detectors. Bottom left: The dependence of
the PDC signal on the pump power.}
\label{setup}
\end{figure}

The experimental setup (Fig~\ref{setup}) contained three main parts. In the state generation part, horizontally polarized third harmonic of a pulsed Nd:YAG laser with the wavelength 354.7 nm, pulse duration 18 ps, and repetition rate 1 kHz was used as a pump. The intensity was changed by rotating a half-wave plate ($\lambda_p/2$) in front of a Glan-Laser Polarizer (GP). The bright two-mode polarization squeezed vacuum state was generated in a frequency-degenerate traveling-wave optical parametric amplifier (OPA) based on two 5-mm width crystals BBO cut for type-II phase matching. The effect of the spatial walk-off was reduced by placing two crystals with the optical axis in the horizontal plane tilted oppositely with respect to the pump. The OPA was tuned to generate twin beams in the collinear direction. A high parametric gain was achieved by reducing the pump beam diameter to $180 ~\mu m$ with the help of a telescope consisting of a convex lens ($F=50$~cm) followed by a concave lens ($F=-7.5$~cm) at a distance of $42.5$~cm. Due to the group velocity difference, the ordinarily polarized pulse of the signal radiation was always delayed relative to the extraordinarily polarized idler pulse. After the crystals the pump was eliminated by two dichroic mirrors $\mathrm {DM_{1,2}}$ with high reflection at $709.3$~nm and high transmission at $354.7$~nm. The rest of the pump was absorbed by the red glass filter RG-630 with AR coating in the spectral range of the PDC radiation. The delay between the signal and idler beams was changed in the delay line with the polarization beamsplitter ($\mathrm {PBS_1}$) at the input by scanning the position of the mirror $\mathrm {M_1}$  with respect to the mirror $\mathrm {M_2}$. The linear polarization of the signal and idler beams was rotated by 90 degrees after double passing through the quarter wave plates ($\lambda_s/4$) oriented at 45 degrees to the $\mathrm {PBS_1}$. Therefore both beams left the interferometer from the same output and were directed to the registration part of the setup. A narrow angular spectrum (with the width $\mathrm {4~mrad}$) 
was restricted by a 3 mm iris aperture (A) placed in the focal plane of a $75$ cm collecting lens ($\mathrm {L_s}$). The HOM interference was observed at the output of the polarization beamsplitter ($\mathrm {PBS_2}$) placed after a halfwave plate ($\lambda_s/2$) which rotated the polarization basis of the $\mathrm {PBS_2}$ by 45 degrees. In this case, one can show that the trajectories of two multi-photon states arriving simultaneously to the polarization beamsplitter became indistinguishable and the interference occurred. After the beamsplitter the radiation was focused by two lenses (L) on the analog detectors ($\mathrm {D_{1,2}}$) based on pin-diodes~\cite{Iskhakov2009}. The electronic pulses from the detectors had a fixed duration ($8 \mu s$) and their amplitudes were proportional to the numbers of photons at the input. Then they were time-integrated with the help of a 10-Msample/s 14-bit analog-to-digital converter card. The card was triggered by the synchronization pulses of the laser. The measured signals per pulse $\mathrm {S_{1,2}}$ at the output of $\mathrm {D_{1,2}}$ were used for the calculation of the normalized second-order intensity correlation function $g^{(2)}=\frac{\langle S_1\cdot S_2\rangle}{\langle S_1\rangle \cdot \langle S_2\rangle}$ and the normalized variance of the difference signal $\frac{\Delta S_-^2}{ \langle S_+\rangle }\equiv\frac{Var(S_1-S_2)}{\langle S_1\rangle + \langle S_2\rangle}$. The angle brackets denote the averaging over an ensemble of pulses. In the experiment, the variance of the difference signal, the second-order intensity correlation function, and the mean values of the signals were calculated for each position of the mirror $\mathrm {M_1}$ by averaging the data over 30000 pulses.

\begin{figure}[htb]
\centering
\includegraphics[width=0.45\textwidth]{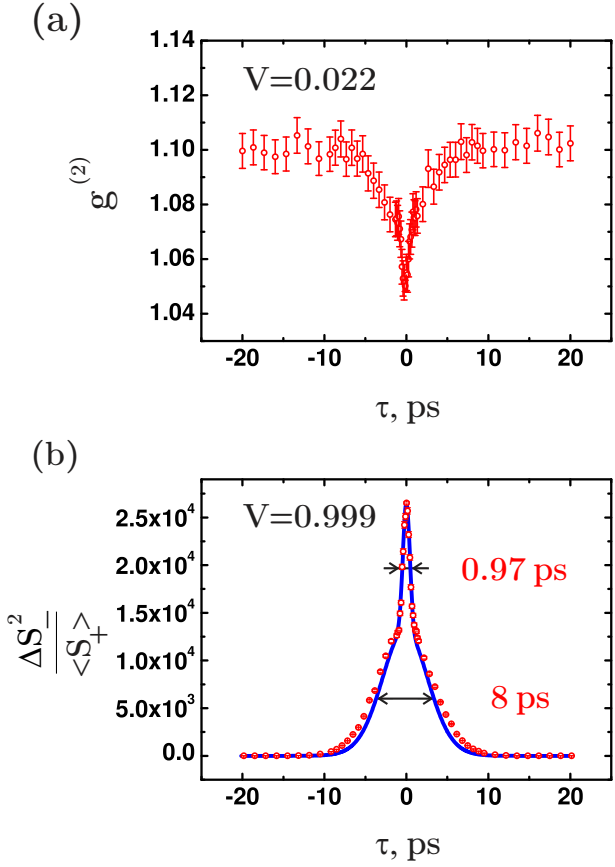}
\caption{(Color online)  HOM interference: dependencies of the second-order intensity correlation function (a) and normalized variance of the difference signal (b) on the delay between signal and idler beams. Red circles represent the experimental results and the blue line is the theoretical dependence.}
\label{res1}
\end{figure}

At the beginning, the brightness of the generated state was evaluated from the parametric gain ($G$) measurement. The intensity of the PDC radiation was measured as a function of the pump power (see bottom-left part of Fig.\ref{setup}). The measurement
was done after narrowband spectral selection ($0.2$ nm around the wavelength of $709.3$ nm) by means of a spectrometer HORIBA Jobin Yvon Micro HR (not shown) and angular selection ($\mathrm {4~mrad}$ around the collinear direction) by means of an aperture. The red circles represent the experimental results and the blue line is the fit plotted according to the dependence $N=\sinh^2G$. The curve clearly demonstrates that the process was strongly nonlinear. The gain reached the value of $G=7.5$ at 55 mW pump power, which is equivalent to $\mathrm{N_{mode}\sim8\cdot10^5}$ photons per mode.
Fig.~\ref{res1} (a) shows the interference dip in the second-order intensity correlation function $g^{(2)}$ plotted versus the delay time. According to the definition of the interference visibility $V\equiv\frac{max-min}{max+min}$, the obtained value of $V=0.022$ was very small. The value of the $g^{(2)}$ at the edges was used for the estimation of the number of detected modes. In this case, the relative delay was large enough and there was no interference at the beamsplitter. Hence, the initial correlation between the signal and idler beams given by the second-order intensity correlation function for a single mode $g^{(2)}=2+\frac{1}{N_{mode}}$ was conserved. Assuming that due to the multi-mode detection the initial normalized second-order intensity correlation function $g^{(2)}$ is transformed into $g_{m}^{(2)}=1+\frac{g^{(2)}-1}{m}$, the number of detected modes was found to be $m=10$.

The normalized variance of the difference signal is displayed as a function of the time delay in Fig.~\ref{res1} (b). Similar to the two-photon HOM interference, multi-photon signal and idler states arriving simultaneously to the beamsplitter were mostly directed as a whole either to one port or to the other. Therefore, the output states of the $\mathrm {PBS_{2}}$ were anticorrelated leading to the observation of a dip in the correlation function or to a peak in the variance of the difference signal. Indeed, a narrow central peak was observed in the experiment. The width of this peak is determined by the coherence time of the PDC radiation. The broader pedestal peak has a classical nature. It results from the pulsed shape of the signal and idler radiation. One should mention here that due to the strong nonlinearity of the parametric process the pulse duration of the PDC radiation was much shorter than the pulse duration of the pump. As it was discussed above, the signal and idler pulses arriving to the beamsplitter in sequence were split independently and their excess fluctuations were eliminated as a result of subtraction. Therefore the value of the normalized variance of the difference signal at the edges corresponds to the shot noise level.
As a result, the peak manifests a visibility $V=0.999$ exceeding all previous records of HOM interference, including the one achieved recently with two oppositely chirped classical optical pulses in an experiment on sum frequency generation~\cite{Resch}.

The theoretical curve shown by the blue line is plotted according to the approach discussed in detail in~\cite{Brambilla2004}. Here we use the Heisenberg picture of a quantum state evolution within the undepleted pump approximation. In this case, the Hamiltonian describing collinear type-II OPA with the signal and idler frequencies $(\pm \Omega)$ tuned around the degenerate frequency $\frac{\omega_p}{2}$ is
\begin{equation}
\hat{H}\propto \Gamma(\Omega)[a_1^{\dag}(\Omega) a^{\dag}_2(-\Omega)]+H.c.,
\label{Ham}
\end{equation}
where $\Gamma(\Omega)$ is the spectral parametric gain coefficient and $a_{1}^{\dag}, a_{2}^{\dag}$ are the photon creation operators in the signal and idler beams. Considering the collinear geometry of the experiment the calculations are done within the plane-wave approximation. The state at the output of the OPA is described using the Bogoliubov transformations~\cite{Leuchs}
\begin{eqnarray}
a_1(\Omega)&=&U(\Omega,t)a^{vac}_1(\Omega)+V(\Omega,t)a^{vac\dag}_2(-\Omega),\nonumber\\
a_2(\Omega)&=&U(\Omega,t)a^{vac}_2(\Omega)+V(\Omega,t)a^{vac\dag}_1(-\Omega),
\label{Bogoliubov}
\end{eqnarray}
where $a_1^{vac},a_2^{vac},a_1^{vac\dag},a_2^{vac\dag}$ are the photon annihilation and creation operators at the vacuum inputs of the OPA. The gain functions $U(\Omega,t)$ and $V(\Omega,t)$ are given by
\begin{equation}
\begin{split}
U(\Omega,t) &= \mathrm{cosh}\Gamma(\Omega,t)+ i\frac{\Delta(\Omega)l_c}{2\Gamma(\Omega,t)}\mathrm{sinh}\Gamma(\Omega,t),\\
V(\Omega,t) &= \frac{G(t)}{\Gamma(\Omega,t)}\mathrm{sinh}\Gamma(\Omega,t)
\end{split}
\end{equation}
with
\begin{equation}
\begin{split}
\Gamma(\Omega,t)=\sqrt{G(t)^2-\frac{\Delta(\Omega)^2l^2_c}{4}},
\end{split}
\end{equation}
where $l_c$ is the total length of the crystals and $G(t)$ depends on the the pump field envelope $A_0(t)$ as $G(t)=\sigma A_0(t)l_c$. Here $\sigma$ is a coefficient depending on the properties of the crystal. As long as the frequency spectrum of type-II PDC is narrow ($\mathrm{1.3~nm}$), the phase mismatch $\Delta$ is defined only by the temporal walk-off between the signal and idler waves at the degenerate frequency $\Delta(\Omega)=[(\partial k_{1}/\partial\omega)_{\omega=\omega_p/2}-(\partial k_{2}/\partial\omega)_{\omega=\omega_p/2}]\Omega$.
Assuming that the signal and idler states arrive at the beamsplitter with the relative delay $\tau$, the beamsplitter transformations take the form
\begin{equation}
a_1^{out}(t)=\frac{a_1(t-\tau)+a_2(t)}{\sqrt{2}},
a_2^{out}(t)=\frac{- a_1(t-\tau)+a_2(t)}{\sqrt{2}}
\end{equation}
Notice that if the pulse duration is $T_p$ the number of photons per pulse at each output of the beamsplitter is $N_{1,2}=\int_{-T_p/2}^{T_p/2}a^{out\dag}_{1,2}(t)a^{out}_{1,2}(t)dt$. According to the definition of the normalized variance of the photon-number difference $\frac{(\Delta N_-)^2}{N_+} \equiv \frac{\langle (N_1-N_2)^2\rangle-\langle N_1-N_2\rangle^2}{\langle N_1+N_2 \rangle}$, with the angle brackets denoting the averaging over the vacuum state, we obtain the expression for the numerical evaluation,

\begin{equation}
\begin{split}
\frac{\Delta N_-^2(\tau)}{\langle N_+\rangle}=1 &+
\frac{1}{\int_{0}^{\infty} d\Omega |V(0,\Omega)|^2}\times\\
\int_{0}^{\infty}d\Omega |V(0,\Omega)|^2 \{ |V(\tau,\Omega)|^2 &+ \hbox{Re}([U^*(0,\Omega)]^2 e^{2i\Omega\tau} ) \}.
\label{NRF}
\end{split}
\end{equation}
Taking into account the finite quantum efficiency $\eta$ of the detection, the actual normalized variance of the difference signal is
\begin{equation}
\frac{\Delta S_-^2(\tau)}{\langle S_+\rangle} = 1+\eta(\frac{\Delta N_-^2(\tau)}{\langle N_+\rangle}-1).
\end{equation}
In accordance with the experiment, numerical calculations were done for two type-II BBO crystals with the total length of $l_c=10~mm$, assuming that the pump pulse has a Gaussian envelope with the duration $\mathrm{T_p=18~ps}$, and a parametric gain of $G=7.5$. The obtained value of the fitting parameter is $\eta=0.03$. The result can be explained by the losses in the optical channel and by the mismatch of the signal and idler beams. This mismatch is due to the fact that although signal and idler beams have different angular spectra because of the anisotropy, they are restricted by the same aperture $\mathrm {A}$. 
Therefore, we find that the theoretical estimations are in a good agreement with the experiment. It is also worth mentioning that the central narrow peak is twice higher then the broad one and its amplitude is twice the mean photon number per mode $2\cdot N$. Since the background is given by a unity, we conclude that the brighter the state the higher visibility can be obtained. The ratio between the widths of the broad pedestal and the central narrow peak gives the effective number of the longitudinal modes, $m_{long}=8$. It is important to mention here that for pure states this ratio is an estimate for the degree of entanglement in the time/frequency domain~\cite{Eberly2002}.

\begin{figure}[htb]
\centering
\includegraphics[width=0.40\textwidth]{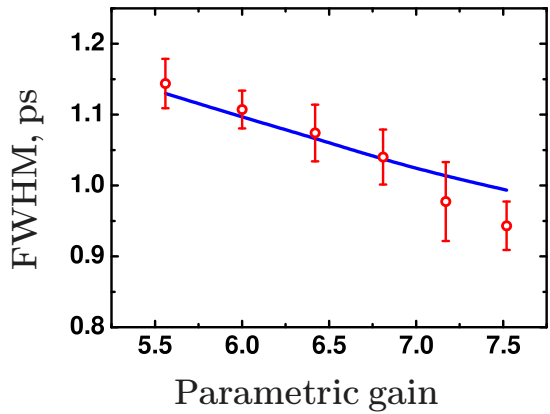}
\caption{(Color online) Full width at half maximum (FWHM) of the HOM peak as a function of the parametric gain. Red circles represent the experimental results and the blue curve is plotted according to the theoretical model.}
\label{width}
\end{figure}

Figure \ref{width} shows the full width at half maximum (FWHM) of the central peak as a function of the parametric gain. A 20\% narrowing of the peak was observed by changing the parametric gain in the range of $5.5<G<7.5$. The red circles are the experimental data, in perfect agreement with the theoretical prediction (the blue line). As it was shown in~\cite{OE2012}, at high gain most of the photons are generated at the output part of the crystal, which leads to the broadening of the spectrum and therefore, to the reduction of the coherence time of the PDC radiation. This makes the experiment more sensitive to the balancing of the arrival times of two multi-photon states and therefore significantly increases the accuracy of the time measurement.

To summarize, we have observed the Hong-Ou-Mandel interference of macroscopic states containing on the average $8\cdot10^5$ photons per mode. By measuring the normalized variance of the difference signal we were able to achieve the visibility of 0.999. This is an evident advantage over the correlation function measurement where the visibility is strongly reduced in the presence of high gain and many modes. The experimental results provide the information on the temporal mode structure of the interfering fields. Moreover, we believe that this measurement can be used for the estimation of the degree of frequency/time entanglement for pure states. It is shown that at high gain PDC the width of the interference peak reduces with the parametric gain leading to an increase in the time resolution. Therefore, we can conclude that the HOM effect for the bright states generated via high-gain PDC has the same properties as the HOM effect for single photons but a better resolution.

We acknowledge the partial support of the Russian Foundation for Basic Research, grants \#10-02-00202, 11-02-01074, and 12-02-00965. T.~Sh.~I. acknowledges funding from Alexander von Humboldt Foundation, and K.~Yu.~S. acknowledges
support from the Dynasty Foundation.

\end{document}